\documentclass[aps,prl,twocolumn,superscriptaddress,showpacs,preprintnumbers,amsmath,amssymb]{revtex4}

\usepackage{graphicx} 
\usepackage{dcolumn}  

\graphicspath{{ps}}

\begin{document}

\preprint{\vbox{ \hbox{BELLE  Preprint 2007-16}
                 \hbox{KEK Preprint 2007-1}
}}

\title{ \quad\\[0.5cm]  \boldmath Search for $B^0 \to p\overline{p}$, $\Lambda \overline{\Lambda}$ and $B^+\to p \overline{\Lambda}$ at Belle}

\affiliation{Budker Institute of Nuclear Physics, Novosibirsk}
\affiliation{Chiba University, Chiba}
\affiliation{University of Cincinnati, Cincinnati, Ohio 45221}
\affiliation{Department of Physics, Fu Jen Catholic University, Taipei}
\affiliation{Hanyang University, Seoul}
\affiliation{University of Hawaii, Honolulu, Hawaii 96822}
\affiliation{High Energy Accelerator Research Organization (KEK), Tsukuba}
\affiliation{Institute of High Energy Physics, Vienna}
\affiliation{Institute of High Energy Physics, Protvino}
\affiliation{Institute for Theoretical and Experimental Physics, Moscow}
\affiliation{J. Stefan Institute, Ljubljana}
\affiliation{Kanagawa University, Yokohama}
\affiliation{Korea University, Seoul}
\affiliation{Kyungpook National University, Taegu}
\affiliation{Swiss Federal Institute of Technology of Lausanne, EPFL, Lausanne}
\affiliation{University of Ljubljana, Ljubljana}
\affiliation{University of Maribor, Maribor}
\affiliation{University of Melbourne, School of Physics, Victoria 3010}
\affiliation{Nagoya University, Nagoya}
\affiliation{Nara Women's University, Nara}
\affiliation{National Central University, Chung-li}
\affiliation{National United University, Miao Li}
\affiliation{Department of Physics, National Taiwan University, Taipei}
\affiliation{H. Niewodniczanski Institute of Nuclear Physics, Krakow}
\affiliation{Nippon Dental University, Niigata}
\affiliation{Niigata University, Niigata}
\affiliation{Osaka City University, Osaka}
\affiliation{Osaka University, Osaka}
\affiliation{Panjab University, Chandigarh}
\affiliation{Peking University, Beijing}
\affiliation{RIKEN BNL Research Center, Upton, New York 11973}
\affiliation{Seoul National University, Seoul}
\affiliation{Shinshu University, Nagano}
\affiliation{Sungkyunkwan University, Suwon}
\affiliation{University of Sydney, Sydney, New South Wales}
\affiliation{Toho University, Funabashi}
\affiliation{Tohoku Gakuin University, Tagajo}
\affiliation{Tohoku University, Sendai}
\affiliation{Department of Physics, University of Tokyo, Tokyo}
\affiliation{Tokyo Institute of Technology, Tokyo}
\affiliation{Tokyo Metropolitan University, Tokyo}
\affiliation{Tokyo University of Agriculture and Technology, Tokyo}
\affiliation{Virginia Polytechnic Institute and State University, Blacksburg, Virginia 24061}
\affiliation{Yonsei University, Seoul}
  \author{Y.-T.~Tsai}\affiliation{Department of Physics, National Taiwan University, Taipei} 
  \author{P.~Chang}\affiliation{Department of Physics, National Taiwan University, Taipei} 
  \author{K.~Abe}\affiliation{High Energy Accelerator Research Organization (KEK), Tsukuba} 
  \author{I.~Adachi}\affiliation{High Energy Accelerator Research Organization (KEK), Tsukuba} 
  \author{H.~Aihara}\affiliation{Department of Physics, University of Tokyo, Tokyo} 
  \author{D.~Anipko}\affiliation{Budker Institute of Nuclear Physics, Novosibirsk} 
  \author{A.~M.~Bakich}\affiliation{University of Sydney, Sydney, New South Wales} 
  \author{U.~Bitenc}\affiliation{J. Stefan Institute, Ljubljana} 
  \author{I.~Bizjak}\affiliation{J. Stefan Institute, Ljubljana} 
  \author{S.~Blyth}\affiliation{National Central University, Chung-li} 
% ...
\collaboration{The Belle Collaboration}

\begin{abstract}

We report on a new search for two-body baryonic decays of the $B$ meson. 
Improved sensitivity compared to previous Belle results is obtained from 414 
fb$^{-1}$ of data that corresponds to 449 million $B \overline{B}$ pairs, 
which were 
taken on the $\Upsilon$(4S) resonance  and collected with the Belle 
detector at the KEKB $e^+e^-$ collider. 
No significant signals are observed and we set the 90\% confidence level 
upper limits: ${\mathcal B}(B^0 \to p \overline p) <1.1\times 10^{-7},
{\mathcal B}(B^0 \to \Lambda \overline \Lambda) < 3.2\times 10^{-7}$ and
${\mathcal B}(B^+ \to p \overline \Lambda) < 3.2\times 10^{-7}$.

\end{abstract}

\maketitle

\tighten

{\renewcommand{\thefootnote}{\fnsymbol{footnote}}}
\setcounter{footnote}{0}


Recent observations of $B$ meson 
decays into several charmless three-body baryonic final states 
show  branching fractions around 10$^{-6}$ and a peak
in the baryon-antibaryon mass spectra near threshold ~\cite{belle-threebody}. 
Further investigations of the angular correlations for events in the threshold
region ~\cite{angle} favor a fragmentation model ~\cite{model}. In contrast, 
charmless two-body baryonic $B$ decays are expected to have lower
branching fractions. However, it is challenging to perform conclusive 
theoretical calculations of  baryon formation in $B$ decay.   
Previous searches \cite{previous-result,current-babar}
for two-body decays yielded upper limits on the 
branching fractions of $ (3-7)\times 10^{-7}$, which are consistent with a 
recent calculation using the pole model~\cite{HYCheng} but in contradiction 
with a calculation 
based on QCD sum rules~\cite{sum-rule-model}. Moreover, the  upper 
limit for $B^0\to p \overline p$ is  also consistent with simple scaling of the
measured branching fraction for $B^0\to \Lambda_c^- p$~\cite{belle-lambda-pbar} 
by the current estimate of $|V_{ub}/V_{cb}|^2$, which gives $ {\cal B}(B^0\to p\overline p) \sim 2.0\times 10^{-7}$. 

Two-body baryonic
decays provide valuable guidance to improve the understanding of quark/gluon 
fragmentation of the $B$ meson to baryons.  In this article we report 
searches with better sensitivity for the charmless two-body baryonic $B$ decays $B^0 \to p\overline{p}, 
B^0 \to \Lambda\overline{\Lambda}$, and $B^+ \to p\overline{\Lambda}$~\cite{charge-conjugate1}.
The analysis is  based on a 414 fb$^{-1}$ data sample, corresponding to 
449 $\times 10^6$ $B\overline{B}$ pairs, accumulated at the $\Upsilon(4S)$ resonance
 with the Belle detector at the KEKB~\cite{KEKB} asymmetric $e^+e^-$ collider.

 
The Belle detector is a large-solid-angle magnetic spectrometer that
consists of a  silicon vertex detector (SVD), a 50-layer
central drift chamber (CDC), an array of aerogel threshold Cherenkov
counters (ACC), time-of-flight scintillation counters (TOF), and an
array of CsI(Tl) crystals, all located inside a superconducting solenoid coil that provides a 1.5 T
magnetic field.  An iron flux-return located outside of the coil is instrumented to detect $K_L^0$ mesons and to identify muons.
The detector is described in detail elsewhere~\cite{Belle}.  
Two different inner detector configurations were used. For the first sample
of 152 million $B\overline B$ pairs, a 2.0 cm radius beampipe
and a 3-layer silicon vertex detector (SVD1) were used;
for the latter  297 million $B\overline B$ pairs,
a 1.5 cm radius beampipe, a 4-layer silicon detector (SVD2) \cite{svd2}
and a small-cell inner drift chamber were used.

Primary charged tracks associated with  candidate $B$ decays are required to satisfy the following criteria: 
the track impact parameters relative to the run-by-run interaction point (IP) are required to be within 
$\pm$2 cm along the $z$ axis (oriented antiparallel to the positron beam)
and within $\pm$0.05 cm in the transverse ($xy$) plane.

Proton candidates are selected based on the likelihood functions $L_p$, $L_K$ 
and $L_{\pi}$ for protons, kaons,
and pions, respectively, which are determined from particle identification information
from the CDC ($dE/dx$ specific ionization), the ACC (Cherenkov radiation pulse height),
and the TOF (time relative to the beam bunch crossing). Charged tracks with
$L_p/(L_p+L_{\pi})>0.6$ are identified as protons and 
$L_{\pi}/(L_p+L_{\pi})>0.6$ as pions. For proton candidates that originate 
directly from $B$ 
decays, an additional requirement, $L_p/(L_p+L_K)>0.6$, is applied to improve 
the purity. The proton identification efficiency  with the tighter requirements 
is 77\% for 2 GeV/$c$ protons.        

$\Lambda$ candidates are reconstructed from oppositely charged pion-proton 
pairs,
satisfying the following criteria:  the
separation distance  between the pion and proton at the $\Lambda$ decay
vertex must be less than 12.9 cm along the $z$ axis; the distance of the 
closest approach in the $xy$ plane  to the IP is greater than  
0.008 cm for each track; the flight length of the 
$\Lambda$ candidates must be greater than 0.22 cm  in the $xy$ plane; the 
angular difference between the $\Lambda$ momentum vector and 
the vector between the IP and the $\Lambda$ decay vertex must be less than
0.09 rad. Finally, 
the reconstructed invariant mass of $\Lambda$ candidate is required to be 
within the mass interval (1.116 $\pm$ 0.005) GeV/$c^2$, corresponding to  
$\pm 3\sigma$ in the mass resolution.



$B$ candidates are reconstructed by pairing two baryons and are identified
with two kinematic variables:
the beam-energy constrained mass, 
$M_{\rm bc}=\sqrt{E_{\rm beam}^2/c^4-p_B^2/c^2}$, 
and the energy difference, $\Delta E=E_B-E_{\rm beam}$, where $E_{\rm beam}$
is the run-dependent beam energy in the center-of-mass (c.m.) frame and
$p_B$ and $E_B$ are reconstructed momentum and energy  of the 
reconstructed $B$ candidates in the c.m. frame, respectively.
The signal region is defined as 
5.27 GeV/$c^2 < M_{\rm bc} <$ 5.29 GeV/$c^2$ and $|\Delta E| <$ 0.05 GeV, while the sideband region is 
defined as  5.2 GeV/$c^2 < M_{\rm bc} <$ 5.26 GeV/$c^2$ and $|\Delta E| <$ 0.2 GeV.

All possible backgrounds are investigated using a GEANT3 ~\cite{geant}
 based Monte Carlo (MC) simulation. Backgrounds from the charmful and 
charmless 
$B$ decays are found to be negligible. The dominant background is from 
continuum  $e^+e^- \to q\overline{q}\;(q=u,d,s,c)$ events, which  are 
studied using the sideband data.

Continuum background is suppressed by requiring 
$|\cos\theta_{\rm th}|<0.9$, where $\theta_{\rm th}$ is the angle in the c.m. frame between the direction of one $B$ daughter and the thrust 
axis~\cite{thrust}, formed by the particles not associated
with the $B$ candidate. The $\cos\theta_{\rm th}$ distribution is nearly flat 
for signal but 
strongly peaks at $\pm 1$ for the continuum.  

\begin{figure}[!htb]
\begin{center}
\resizebox*{3.5in}{3.in}{\includegraphics{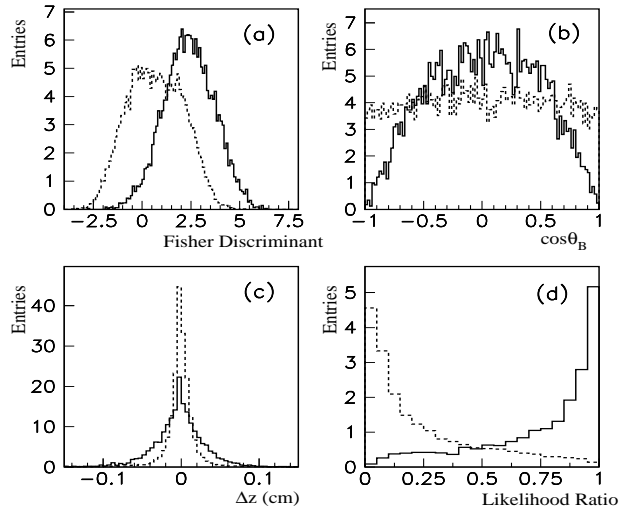}}
\end{center}
\caption{Distributions of (a) the  Fisher discriminant, (b) $\cos\theta_B$, 
(c) $\Delta$z, and (d) likelihood ratio for $B^0\to p\overline p$ candidates . 
The solid histogram is for signal 
MC, while the dashed histogram is for sideband events.}
\label{fig:shape}
\end{figure}

The background rejection is further improved  using  
event topology, $B$ candidate vertex and $B$ flavor tagging information. 
First, we combine a set of modified Fox-Wolfram moments \cite{pi0pi0} into a
Fisher discriminant~\cite{fisher} ${\mathcal F}$ to distinguish  spherically distributed 
$B\overline{B}$ events from the jet-like continuum backgrounds. Figure
\ref{fig:shape}(a) shows the Fisher discriminant for $p\overline p$ signal MC 
and sideband  events. We then use the cosine of the  angle $\theta_B$ between 
the $B$ candidate flight direction
and the $z$ axis (Fig.~\ref{fig:shape}(b)).  For the $p\overline{p}$ mode only,
 we use the distance
$\Delta z$ along the $z$ axis between the $p\overline{p}$ vertex and the vertex 
formed from the
remaining charged tracks (Fig. \ref{fig:shape}(c)).
The $\Delta z$ probability density function is modeled by a double 
Gaussian independently
for the SVD1 and SVD2 data due to the better vertex resolution provided by the 
SVD2. The PDFs for $\cos\theta_B$ and ${\mathcal F}$ are described by a 2nd
order polynomial and a double asymmetric Gaussian, respectively.
The quantities ${\mathcal F}$, $\cos\theta_B$ and $\Delta z$ (for the
$p \overline p$ mode only)  are combined to form likelihoods ${\mathcal L}_S$ and ${\mathcal L}_B$ for
signal and continuum background, respectively.  
The normalized ratio ${\mathcal R = \mathcal L_S/(\mathcal L_S+\mathcal L_B)}$,
shown in Fig.  \ref{fig:shape}(d), peaks at unity for signal and at zero for continuum.


The distribution of ${\mathcal R}$ is somewhat correlated with the event's $B$
flavor information.  The standard Belle flavor tagging algorithm \cite{tagging}
provides a tagging quality factor $r$ that
ranges from zero for no flavor tagging information to unity for unambiguous flavor assignment. 
 We apply a requirement on the likelihood ratio ${\mathcal R}$ depending on the
value of $r$ (and on the inner detector configuration for the 
$p{\overline p}$ mode).
The likelihood ratio requirements listed in Table~\ref{table:LR_Cut}
are determined by optimizing the figure-of-merit, 
$N_S/\sqrt{N_S+N_B}$, where $N_S$ and $N_B$ are the expected signal and 
background yields. The signal yields are estimated from the product of the 
number of 
$B\overline B$ events, the signal efficiency from  MC and the assumed 
branching fraction of $10^{-7}$. The expected  background yields are obtained
by scaling the amount of data in the region, 
$|\Delta E| < 0.05\,{\rm GeV}$ and
$5.2\,{\rm GeV}/c^2 < M_{\rm bc} < 5.26\,{\rm GeV}/c^2$,
 to the signal region.
The $M_{\rm bc}$ distribution of the continuum is assumed
to be independent of $\Delta E$. Therefore, we model 
the $M_{\rm bc}$ distribution with an ARGUS
function~\cite{argus}  using  the data in 0.1 GeV $< |\Delta E| <$ 0.2 GeV.
The scaling factor is estimated
to be the ratio of the area of the $M_{\rm bc}$ signal region to the sideband
region based on the ARGUS function.

\begin{table}[htpb]
\begin{center}
\caption{Minimum values of the likelihood ratio ${\mathcal R}$ for three ranges
   of the flavor tagging quality $r$ 
(and for the inner detector configuration for the $p\overline{p}$ mode).}
\label{table:LR_Cut}
\vspace*{0.2cm}
\begin{tabular}{ccccccc}
\hline
\hline
    &\multicolumn{2}{c}{$\; 0 \le r < 0.5 \;\;\;$}  & \multicolumn{2}{c}{$\; 0.5 \le r \le 0.75\;$} &\multicolumn{2}{c}{
 $\; 0.75 < r \le 1.0\;$} \\
    & SVD1 & SVD2 & SVD1 & SVD2 & SVD1 & SVD2 \\
\hline
 B$^0 \to p\overline{p}$               & 0.85 & 0.85 & 0.8 & 0.65 & 0.65 & 0.65 \\
 B$^0 \to \Lambda\overline{\Lambda}$   & \multicolumn{2}{c}{0.85} & \multicolumn{2}{c}{0.6} & \multicolumn{2}{c}{0.35} \\
 B$^+ \to p\overline{\Lambda}$         & \multicolumn{2}{c}{0.8} & \multicolumn{2}{c}{0.75} & \multicolumn{2}{c}{0.65} \\
\hline
\hline
\end{tabular}
\end{center}
\end{table}

The $B$ signal efficiency for each mode is obtained 
using the MC simulation after 
applying all analysis requirements except the proton identification. The $B$ 
 signal efficiency is  corrected for the proton identification; the proton 
 identification  efficiency is estimated  
using $\Lambda\to p\pi^-$ decays in  data. 
The systematic error in the $B$ signal efficiency arises from 
proton identification,
 tracking efficiency,  $\Lambda$ selection and the likelihood ratio requirement
in each flavor-tagged region.  The statistical error in the proton efficiency 
obtained in the $\Lambda$ sample is included in the systematic error for 
proton identification.
The tracking systematic error is studied using
an inclusive $D^{*+}\to D^0\pi^+$ sample, where $D^0\to K^0_S \pi^+\pi^-$ and
$K^0_S\to \pi^+\pi^-$. Candidate $D^{*+}$ mesons are kinematically identified 
using the momentum of the slow pion  from the $D^{*+}$ decay, 
 $D^0$ and $K_S^0$ mass constraints, and the trajectory of one
of the pions in the $K^0_S$ decay. The tracking efficiency is measured by
searching for the other $K^0_S$ daughter; by comparing the efficiency in data 
with the Monte Carlo expectation, the tracking  systematic error is
found to be 1\% per track for $p> 0.3$ GeV/$c$ but slightly larger for low
momentum particles.  The systematic error associated with requiring a 
detached vertex for the $\Lambda$ candidate is studied by comparing the
ratio of $D^+\to K_S^0\pi^+$ and $D^+\to K^-\pi^+\pi^+$ yields
with the MC expectation. The resulting $K_S^0$ detection systematic
error is $\pm 4.5\%$, which is used as the $\Lambda$ systematic error.
The likelihood ratio requirement is studied using 
control samples with the same number of charged particles in the
final state:  $B^0 \to K^+\pi^-$ for $p\overline{p}$, 
$B^0 \to D^-\pi^+ \to (K^+\pi^-\pi^-)\pi^+$ for $\Lambda\overline{\Lambda}$ and
 $B^+ \to \overline{D}^0\pi^+ \to (K^+\pi^-)\pi^+$ for $p\overline{\Lambda}$. 
The total systematic uncertainty is computed by adding the 
correlated errors linearly, and then adding the uncorrelated ones in quadrature.
The systematic errors  are summarized in Table ~\ref{table:systematic_errors}.


\begin{table}[htpb]
\begin{center}
\caption{Summary of systematic errors, given in \%.}
\label{table:systematic_errors}
\vspace*{0.2cm}
\begin{tabular}{cccc}
\hline
\hline 
    & $\;\;\; B^0 \to p\overline{p}\;\;\;$  & $\;\;\; B^0 \to \Lambda\overline{\Lambda}\;\;\;$  & $\;\;\; B^+ \to p\overline{\Lambda}\;\;\;$ \\
\hline
 Tracking               & 2.00   & 4.29   & 3.16    \\
 PID                  & 0.40 & 0.12   & 0.26    \\
 $\mathcal R$ Requirement               & 3.70 & 1.22   & 0.75    \\
 $\Lambda$ Selection  & $-$   & 4.50$\times$ 2    & 4.50  \\
 ${\cal B}(\Lambda \to p\pi^-) $ & $-$ & $1.56$ & $0.78$ \\
 \# of $B\overline B$       & 1.27  & 1.27    & 1.27     \\ 
\hline
 Total ($\%$)               & 4.42 & 10.24   & 5.76    \\
\hline
\hline
\end{tabular}
\end{center}
\end{table}





The expected background contribution is obtained using the sideband region 
 described above.  
A  loose $\mathcal{R}$ requirement is applied to ensure large  statistics
to determine the ARGUS parameters.    
The systematic uncertainty in the backround prediction  comes from the
fit errors of the ARGUS parameters and the statistical errors of the events
in the sideband region.



The estimated background yields in the signal region are listed in
Table~\ref{tab:eff} and  are close to the number of observed events.
Since no statistically significant signals are found, we follow the
Feldman-Cousins approach~\cite{F.C.} to estimate 90\% confidence level 
(C.L.) upper limits on the signal yields, using the implementation of J.~Conrad 
\emph{et al.} \cite{pole} to include the systematic errors.
The final results are listed in Table~\ref{tab:eff}.

\begin{figure}[!htb]
\begin{center}
\resizebox*{3.5in}{3.2in}{\includegraphics{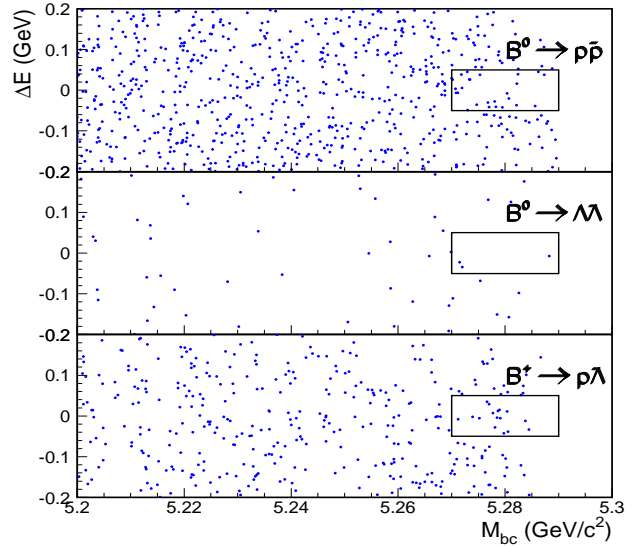}}
\end{center}
\caption{Scatter plots of $\Delta E$ {\rm vs.} $M_{\rm bc}$ for 
$B^0 \to p\overline{p}$ (upper), $B^0 \to \Lambda\overline{\Lambda}$ (middle), and $B^+ \to p\overline\Lambda$ (lower). The
   signal region is indicated by the rectangle in each plot.}
\label{fig:2d}
\end{figure}     

\begin{table}[]
\begin{footnotesize}
\caption{Summary of the $B^0 \to p \overline{p},\, \Lambda \overline{\Lambda}$ and 
$B^+ \to p \overline{\Lambda}$ searches, where $\epsilon$ is the reconstruction efficiency including the sub-decay branching fraction, 
$N_{\rm obs}$ is the observed number of events in the signal region, 
$N^{\rm bg}_{\rm exp}$ is the expected background in the signal region,
$N_{\rm 90}$ is the  yield limit at the 90\% confidence level, and BF 
is  corresponding upper 
limit for the branching fraction. The  
uncertainty in $N^{\rm bg}_{\rm exp}$ is the 
systematic uncertainty due to the ARGUS parameters and statistical error
of the sideband sample. }
\label{tab:eff}
\begin{center}
\begin{tabular}{lccccc}
\hline
\hline
Mode & \hspace{.5cm} $\epsilon$ [\%] & $\;\;\; N_{\rm obs}\;\;\;$ & $\;\;\;N^{
\rm bg}_{\rm exp}\;\;\;$ & $\;\;\; N_{\rm 90} \;\;\;$ & 
\hspace{0.3cm} BF [$10^{-7}$]\\
\hline
$B^0 \to p \overline{p}$ &\hspace{0.3cm} 17.73 & 25 & 26.8$\pm$2.3 &   8.7 &$< 1.1$ \\
$B^0 \to \Lambda \overline{\Lambda}$ & \hspace{0.3cm} 4.24 & 3 & 1.2$\pm$0.5 & 
6.1 & $< 3.2$ \\
$B^+ \to p \overline{\Lambda}$ &\hspace{0.3cm} 8.44 & 16 & 12.9$\pm$1.7 & 
 12.1 & $< 3.2$ \\
\hline \hline
\end{tabular}
\end{center}
\end{footnotesize}
\end{table}                         

In summary, we have performed a search for the decays 
$B^0 \to p\overline{p}, \Lambda\overline{\Lambda}$, and $B^+ \to p\overline{\Lambda}$ 
in a sample of $449 \times 10^6$ $B\overline{B}$ events, which is three times 
larger than the dataset used in the 
previous analysis~\cite{previous-result}. We find no evidence for signals 
and place 90\% C.L. upper limits on the branching fractions of 
$1.1 \times 10^{-7}$, 
$3.2 \times 10^{-7}$, and $3.2 \times 10^{-7}$ for the $p\overline{p}$, 
$\Lambda\overline{\Lambda}$, and $p\overline{\Lambda}$ modes, respectively.
These upper limits improve our previous 
results~\cite{previous-result} and are more stringent than other experimental 
limits~\cite{current-cleo, current-babar}. Moreover, although our 
$p\overline p$ and 
$p\overline{\Lambda}$ results already reach
the pole model predictions~\cite{HYCheng}, no clear signals are seen.

We thank the KEKB group for the excellent operation of the
accelerator, the KEK cryogenics group for the efficient
operation of the solenoid, and the KEK computer group and
the National Institute of Informatics for valuable computing
and Super-SINET network support. We acknowledge support from
the Ministry of Education, Culture, Sports, Science, and
Technology of Japan and the Japan Society for the Promotion
of Science; the Australian Research Council and the
Australian Department of Education, Science and Training;
the National Science Foundation of China and the Knowledge
Innovation Program of the Chinese Academy of Sciences under
contract No.~10575109 and IHEP-U-503; the Department of
Science and Technology of India; 
the BK21 program of the Ministry of Education of Korea, 
the CHEP SRC program and Basic Research program 
(grant No.~R01-2005-000-10089-0) of the Korea Science and
Engineering Foundation, and the Pure Basic Research Group 
program of the Korea Research Foundation; 
the Polish State Committee for Scientific Research; 
the Ministry of Education and Science of the Russian
Federation and the Russian Federal Agency for Atomic Energy;
the Slovenian Research Agency;  the Swiss
National Science Foundation; the National Science Council
and the Ministry of Education of Taiwan; and the U.S.\
Department of Energy.



\begin{thebibliography}{99}

\bibitem{belle-threebody} M.-Z. Wang {\it et al.} (Belle Collaboration), Phys. Rev. Lett. {\bf 90}, 201802 (2003); M.-Z. Wang {\it et al.} (Belle Collaboration), Phys. Rev. Lett. {\bf 92}, 131801 (2004); Y.J. Lee {\it et al.} 
(Belle Collaboration) Phys. Rev. Lett. {\bf 95}, 061802 (2005).

\bibitem{angle}  M.-Z. Wang {\it et al.} (Belle Collaboration), Phys. Lett. 
B {\bf 617}, 141 (2005). 

\bibitem{model} J.L. Rosner, Phys. Rev. D {\bf 68}, 014004 (2003).   


\bibitem{previous-result} M.-C. Chang {\it et al.} (Belle Collaboration), Phys. Rev. D {\bf 71}, 072007 (2005).

\bibitem{current-babar} B.~Aubert {\it et al.} (BABAR Collaboration), Phys. Rev. D {\bf 69}, 091503 (2004).

\bibitem{HYCheng} H.Y. Cheng and K.C. Yang, Phys. Rev. D {\bf 66}, 014020 (2002).

\bibitem{sum-rule-model} V.L. Chernyak and I.R. Zhitnitsky, Nucl. Phys. B {\bf 345}, 137 (1990); this sum-rule calculation predicts that ${\cal B}(B^0\to p\overline p) = 3\times 10^{-7}$ and the branching fractions of charmless two-body
baryonic decays through $b\to s$ penguin transition are $(0.3\sim 1.0)\times 10^{-5}$.

\bibitem{belle-lambda-pbar} N.~Gabyshev {\it et al.} (Belle Collaboration), Phys. Rev. Lett. {\bf 90}, 121802 (2003).

\bibitem{charge-conjugate1} Charge conjugate modes are implicitly included throughout this paper.

\bibitem{KEKB} S.~Kurokawa and E.~Kikutani, Nucl. Instr. and Meth. A {\bf 499},
 1 (2003), and other papers included in this Volume. 

\bibitem{Belle} A. Abashian {\it et al.} (Belle Collaboration), Nucl. Instr. and Meth. A {\bf 479}, 117 (2002).

\bibitem{svd2} Z. Natkaniec {\it et al.} (Belle SVD2 Group), Nucl. Instr.
and Meth. A {\bf 560}, 1 (2006).

\bibitem{geant} R.~Brun {\it et al.}, GEANT 3.21, CERN Report No. DD/EE/84-1 (1987).

\bibitem{thrust} E.~Farhi, Phys. Rev. Lett. {\bf 39}, 1587 (1977).

\bibitem{pi0pi0} 
G.C. Fox and S. Wolfram, Phys. Rev. Lett. {\bf 41}, 1581 (1978).
The modified moments used in this paper are described in, S.H. Lee
{\it et al.} (Belle Collaboration), Phys. Rev. Lett. {\bf 91},
261801 (2003).


\bibitem{fisher}R.A.~Fisher, Ann. Eugenics {\bf 7}, 355 (1937).

\bibitem{tagging} H.~Kakuno {\it et al.} Nucl. Instr. and Meth. A {\bf 533}, 516 (2004).

\bibitem{argus} H.~Albrecht {\it et al.} (ARGUS Collaboration), Phys. Lett. B {\bf 241}, 278 (1990); {\bf 254}, 288 (1991).     

\bibitem{F.C.} G.~J. Feldman and R.~D. Cousins, Phys. Rev. D {\bf 57}, 3873 (1998).

\bibitem{pole} J.~Conrad, O.~Botner, A.~Hallgren and C. Perez de los Heros, Phys. Rev. D {\bf 67}, 012002 (2003). 


\bibitem{current-cleo} A.~Bornheim {\it et al.} (CLEO Collaboration), Phys. Rev. D {\bf 68}, 052002 (2003).






\end{thebibliography}
\end{document}